\documentclass{optica-article}

\journal{opticajournal} 

\articletype{Research Article}

\usepackage{lineno}

\usepackage{graphicx}
\usepackage{dcolumn}
\usepackage{bm}
\usepackage{braket}
\usepackage{amsmath}
\usepackage{bm}
\usepackage{setspace}
\usepackage[euler]{textgreek}
\allowdisplaybreaks
\usepackage{multirow}
\usepackage{comment}
\usepackage{booktabs}
\usepackage{relsize}
\usepackage{hyperref}
\usepackage{siunitx}
\hypersetup{
    colorlinks=true,
    linkcolor=blue,
    filecolor=magenta,      
    urlcolor=cyan,
    citecolor=red
}
\usepackage{upgreek}
\newcommand{\rdown}{r_{\downarrow}}
\newcommand{\rup}{r_{\uparrow}}
\newcommand{\psucc}{p_\text{succ}}
\newcommand{\kwg}{\kappa_\text{wg}}
\newcommand{\lamcav}{\lambda_\text{cav}}
\newcommand{\lamsnv}{\lambda_\text{SnV}}
\newcommand{\etadet}{\eta_{\text{det}}}
\newcommand{\etaexc}{\eta_{\text{exc}}}

\begin{document}

\title{A scalable cavity-based spin-photon interface in a photonic integrated circuit}

\author{Kevin C. Chen\authormark{1,*}, Ian Christen\authormark{1}, Hamza Raniwala\authormark{1}, Marco Colangelo\authormark{1}, Lorenzo De Santis\authormark{1,2}, Katia Shtyrkova\authormark{3}, David Starling\authormark{3}, Ryan Murphy\authormark{3}, Linsen Li\authormark{1}, Karl Berggren\authormark{1}, P. Benjamin Dixon\authormark{3}, Matthew Trusheim\authormark{1,4}, Dirk Englund\authormark{1}}

\address{\authormark{1}Department of Electrical Engineering and Computer Science, Massachusetts Institute of Technology, Cambridge, MA 02139, USA\\
\authormark{2}Currently at QuTech, Delft University of Technology, Delft, Netherlands\\
\authormark{3}Lincoln Laboratory, Massachusetts Institute of Technology, Lexington, MA 02421, USA\\
\authormark{4}DEVCOM, Army Research Laboratory, Adelphi, MD 20783, USA}

\email{\authormark{*}kcchen@mit.edu}


\begin{abstract*} 
A central challenge in quantum networking is transferring quantum states between different physical modalities, such as between flying photonic qubits and stationary quantum memories. One implementation entails using spin-photon interfaces that combine solid-state spin qubits, such as color centers in diamond, with photonic nanostructures. However, while high-fidelity spin-photon interactions have been demonstrated on isolated devices, building practical quantum repeaters requires scaling to large numbers of interfaces yet to be realized. Here, we demonstrate integration of nanophotonic cavities containing tin-vacancy (SnV) centers in a photonic integrated circuit (PIC). Out of a six-channel quantum micro-chiplet (QMC), we find four coupled SnV-cavity devices with an average Purcell factor of $\sim$7. Based on system analyses and numerical simulations, we find with near-term improvements this multiplexed architecture can enable high-fidelity quantum state transfer, paving the way towards building large-scale quantum repeaters.
\end{abstract*}

\section{Introduction}
A principal goal in quantum information science is efficiently distributing entanglement over large distances by transferring quantum states between different physical modalities, such as stationary atomic systems and propagating optical fields. Recent advances in spin-photon interfaces based on color centers in diamond have enabled multi-modality entanglement distribution demonstrations, such as foundational quantum entanglement tests~\cite{Bernien_2013} and construction of a multi-node quantum network~\cite{Pompili_2021}. In particular, the Group-IV centers are promising candidates due to their optical coherence even when placed in nanostructures~\cite{Evans_2016,Wan_2020,Debroux_2021,Martinez_2022_PRL,Parker_2023}, high Debye-Waller factors~\cite{Thiering_2018}, and long spin coherence times~\cite{Sukachev_2017,Debroux_2021}. These aforementioned attributes enable demonstrations of high-fidelity quantum state transfer between spins and photons~\cite{Bhaskar_2020} and a multi-qubit repeater node with error detection~\cite{Stas_2022}. The tin-vacancy (SnV) center in the Group-IV family especially has garnered a surging interest due to its large spin-orbital splitting~\cite{Iwasaki_2017_PRL,Thiering_2018}, which suppresses phonon-induced decoherence and permits quantum operations at temperatures above 1~K~\cite{Debroux_2021}, as opposed to the silicon-vacancy centers that require sub-Kelvin temperatures to attain millisecond coherence times~\cite{Sukachev_2017}. Further photonics engineering around SnV centers has also led to production of highly indistinguishable single photons out of nanophotonic waveguides~\cite{Martinez_2022_PRL} and coherent coupling to photonic crystal cavities~\cite{Rugar_2021,Kuruma_2021}.

However, the outstanding challenge towards useful quantum repeaters lies in scaling to $>$$10^2$ spin-photon interfaces to counteract latency and consequently memory decoherence~\cite{van_Dam_2017,Shchukin_2019,Lee_2020,Dai_2021,Dhara_2021_PRA,Choi_2023}, motivating the integration of color centers with photonic integrated circuits (PICs). This hybrid approach~\cite{Benson_2011,Davanco_2017,Elshaari_2020_qPIC,Kim_2020_qPIC,Moody_2022_qPIC} leverages foundry-scale manufacturing for essential quantum control functions, including quantum emitter strain tuning~\cite{Wan_2020,Clark_2023}, spin ground state control~\cite{Golter_2023}, and in-situ optical routing and addressing~\cite{Wan_2020,Palm_2023}. While coupling quantum emitters to nanocavities has enabled excellent spin-photon cooperativity in single and isolated devices~\cite{Nguyen_2019_PRL}, realizing cavity-enhanced interfaces has thus far remained an open challenge in scalable hybrid quantum PIC approaches.

\begin{figure}
    \centering
    \includegraphics[width=\textwidth]{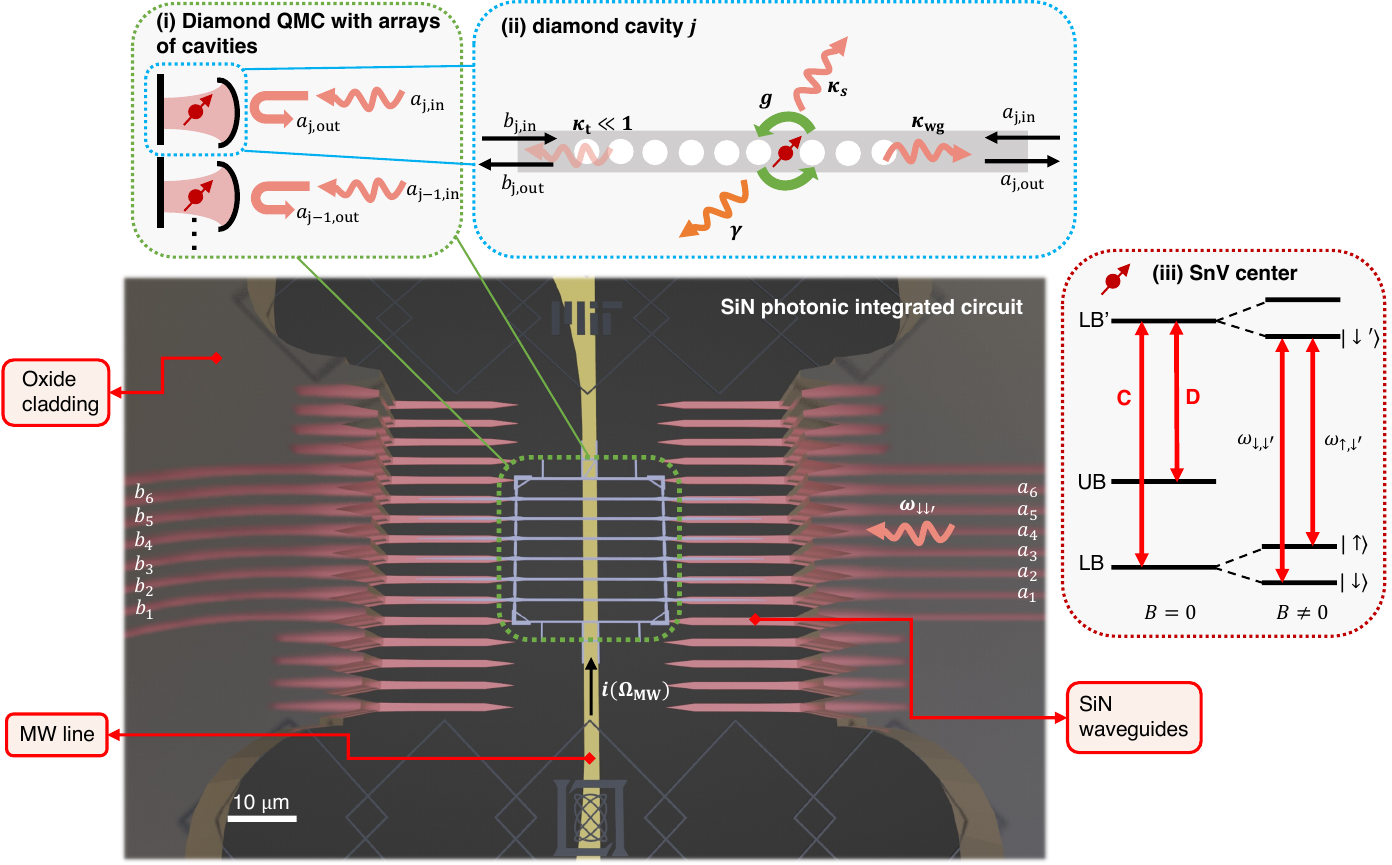}
    \caption{Overview of the hybrid PIC architecture. Oxide-cladded SiN waveguides route from the edge of the chip to an oxide window in which a diamond QMC is placed. Each QMC channel is evanescently coupled to an exposed 100~nm-thick SiN waveguide, and a microwave line (at frequency $\Omega_\text{MW}$) beneath the diamond nanostructure is used to coherently drive the spin qubit. The QMC contains (i) an array of single-sided cavities with input mode $a_{j,\text{in}}$ that enable cavity-reflection-based protocols~\cite{Duan_2004}. Each single-sided cavity is (ii) a 1D photonic crystal cavity in diamond coupled to a SnV center with coupling strength $g$, and the emitter can spontaneously emit into free space at rate $\gamma$. The cavity mode is coupled to both the adjacent waveguide mode  and free space (leakage) at rates $\kwg$ and $\kappa_s$, respectively. The waveguide-cavity interaction can be modeled by the input-output formalism~\cite{Gardiner_Collett_1985} with operators $\{a_{j,\text{in}},a_{j,\text{out}}\}$. The transmission port is assumed to be weakly coupled to the cavity with rate $\kappa_t\ll 1$, with corresponding operators $\{b_{j,\text{in}},b_{j,\text{out}}\}$. (iii) At zero field ($B=0$), the ground state manifold and the lower excited state of SnV center defines two optical transitions, C and D. With an applied field ($B\neq 0$) that gives rise to Zeeman splitting of spin states $\{\ket{\downarrow},\ket{\uparrow}\}$, the C transition further splits into two at frequencies $\{\omega_{\downarrow,\downarrow'},\omega_{\uparrow,\downarrow'}\}$.}
    \label{fig:overview_architecture}
\end{figure}

Here, we address this challenge by introducing SnV centers that are coupled to integrated diamond cavities in a quantum microchiplet (QMC). Figure~\ref{fig:overview_architecture} shows a hybrid architecture that integrates a diamond QMC with a silicon nitride (SiN) PIC for optical and electrical addressing of the individual color centers. The oxide-cladded photonic chip contains an oxide opening, in which individual low-loss nitride waveguide evanescently couples to each diamond QMC channel. Each channel $j$ includes a diamond nanophotonic cavity coupled to input-output waveguide mode $a_j$. In our study, we first demonstrate Purcell enhancement of four cavity-coupled SnV systems out of a 6-channel QMC with an average Purcell factor of $\sim$7. Second, we heterogeneously integrate a separate cavity QMC into a SiN PIC and perform spectroscopy enabled by on-chip coupling. Based on numerical analyses, this quantum PIC indicates that near-term improvements in cavity-PIC coupling can enable multiplexed quantum repeaters for high-fidelity quantum state transfer.

\section{Diamond nanophotonic cavities}

\begin{figure}
    \centering
    \includegraphics[width=\textwidth]{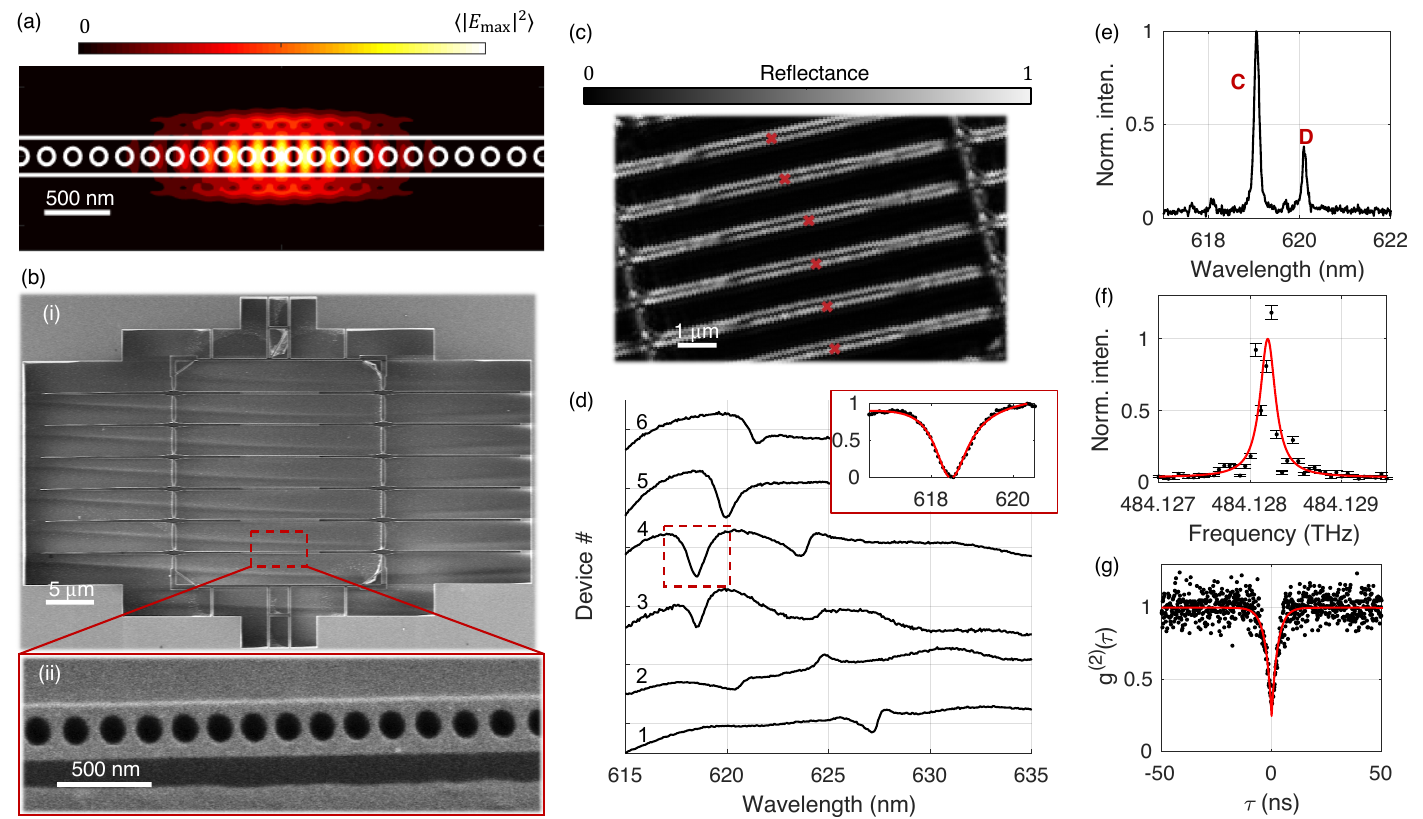}
    \caption{Characterization of diamond PhC cavities and SnV centers. (a) The fundamental mode of the 1D PhC cavity simulated in FDTD. (b) Scanning electron microscope (SEM) images of (i) six fabricated diamond PhC cavities in a QMC and (ii) the center region of cavity 1. (c) A confocal reflection map showing the six cavities and the selected positions for top excitation and collection. (d) The corresponding cavity reflection spectra (room temperature) with Fano lineshapes indicating cavity resonances. Inset: A fundamental resonance of cavity 4, with a fitted $Q=634\pm 14$ at $\lamcav=618.45\pm 0.02~$nm. (e) A representative PL spectrum of a SnV center in cavity 4 showing both C and D transitions at 4~K. (f) A PLE curve of the same SnV center in a cavity region fitted to a Lorentzian, with resonance $\lamsnv=484.12820\pm 0.00004~$THz and linewidth $\sim$$204\pm 71~$MHz. (g) An autocorrelation measurement showing a $g^{(2)}(0)=0.25\pm 0.01$ at zero time delay.}
    \label{fig:diamond_spectroscopy}
\end{figure}

Critical to engineering strong emitter-cavity interaction is maximizing the ratio of cavity quality factor $Q$ to mode volume $V$~\cite{Purcell_1946}. Hence, we design a diamond 1D photonic crystal (PhC)~\cite{Sipahigil_2016,Mouradian_2017,Nguyen_2019_PRL,Rugar_2021,Knall_2022} cavity based on a Gaussian chirp (see Supplement 1, Sec. 1). The fundamental TE-like mode achieves simulated $Q=3\times 10^6$ and $V=0.8(\lambda/n)^3$, with mode profile depicted in Fig.~\ref{fig:diamond_spectroscopy}(a). Using the quasi-isotropic etching technique for undercutting diamond devices~\cite{Khanaliloo_2015,Mouradian_2017,Wan_2018}, we fabricate diamond QMCs containing arrays of 1D PhC cavities (see Supplement 1, Sec. 1). Figure~\ref{fig:diamond_spectroscopy}(b) illustrates representative devices with a measured geometry defined by the width $W=260~$nm, the thickness $H=201~$nm, and the hole radius $r=64~$nm. To characterize the devices, we excite the cavities' modes normal to the plane of the sample (marked in Fig.~\ref{fig:diamond_spectroscopy}(c)) with a pulsed broadband laser and collect its cross-polarized reflection confocally~\cite{Englund_2007}. Due to undercoupling between the cavity and the waveguide modes, i.e. $\kwg\leq 1$, we find top collection of the scattered photons into free-space enables easier characterization of the cavities' spectral responses. Figure~\ref{fig:diamond_spectroscopy}(d) shows examples of cavity reflection spectra acquired at room temperature, with QMC channels 2-6 exhibiting Fano resonances~\cite{Fan_2003} close to 619~nm. We fit these modes to a weighted Fano-Lorentz function~\cite{Avrutsky_2013} (see Supplement 1, Sec. 2). An example fit to channel 4's resonance is shown in the inset of Fig.~\ref{fig:diamond_spectroscopy}(d), giving a fitted resonance wavelength $\lamcav=618.45\pm 0.02~$nm and $Q=(6.34\pm 0.14)\times 10^2$. We notice that the cavity $Q$ improves by a factor of $\sim$2.4 as we cool the sample down to 4~K. We attribute the rest of the discrepancy from the simulated $Q$ to scattering loss due thickness non-uniformity as observed for the underside in Fig.~\ref{fig:diamond_spectroscopy}(b)(ii), which stems from quasi-isotropic etching~\cite{Wan_2018}, as well as surface roughness (see Supplement 1, Sec. 1).

\section{Coherent cavity coupling of quantum emitters}

To investigate coherent emitter-cavity coupling, we perform spectroscopy of SnV centers in the same QMC as shown in Fig.~\ref{fig:diamond_spectroscopy}(c) at 4~K. Figure~\ref{fig:diamond_spectroscopy}(e) shows the photoluminescence spectrum of a SnV center in channel 4 excited off-resonantly at 515~nm. We observe both the C and D transitions at near 619~nm separated by its ground state spin-orbital splitting of $\sim$820~GHz. To probe its optical coherence, we perform photoluminescence excitation (PLE) spectroscopy, in which we sweep a narrowband (linewidth $<50~$kHz) tunable laser across the C transition. Figure~\ref{fig:diamond_spectroscopy}(f) shows the collected phonon sideband (PSB) fluorescence. We find the fitted Lorentzian center and linewidth to be $c/\lamsnv=484.12820(4)~$THz and $\Gamma=204\pm 71~$MHz, which indicates optical dephasing that broadens the linewidth by a factor of $\sim$7 of the transform limit defined by its off-resonance lifetime (see Supplement 1, Sec. 8).

\begin{figure}[h!]
    \centering
    \includegraphics[width=0.5\textwidth]{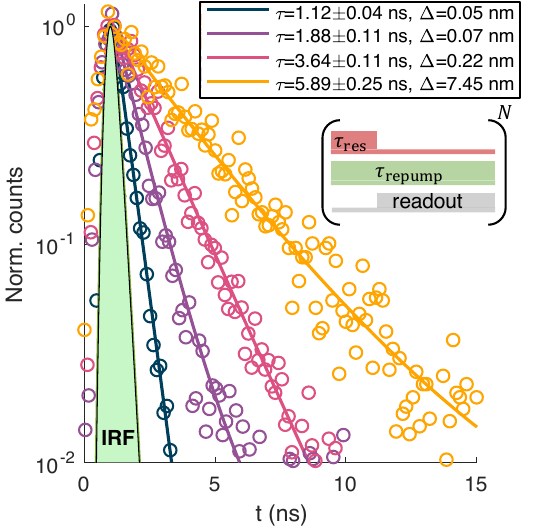}
    \caption{Purcell enhancement of multiple SnV-cavity systems in a single QMC. A representative (channel 6) lifetime $\tau$ versus emitter-cavity detuning $\Delta$ curve is fitted with a convolution between a single exponential and a Gaussian that represents the IRF (shaded in green). The emitter's lifetime reduces as detuning decreases due to Purcell enhancement. Each pulse sequence (repeated $N$ times over 200~s) consists of a short resonant pulse with weak CW repump light, followed by readout of the PSB fluorescence. $t=0$ indicates the beginning of the instrument response function (IRF).}
    \label{fig:purcell}
\end{figure}

Subsequently, we perform an autocorrelation $g^{(2)}$ measurement with a Hanbury-Brown-Twiss setup by resonantly exciting the SnV center. Figure~\ref{fig:diamond_spectroscopy}(g) shows a histogram of correlated photon counts, with $g^{(2)}(\tau)$ plotted against the time delay $\tau$. We fit the data to the model $g^{(2)}(\tau)\propto 1-\exp\left(-|\tau|/\tau_0\right)$, where $\tau_0=2.74\pm 0.17$~ns is a convolved time scale between the emitter's lifetime and the resonantly-driven Rabi oscillation rate~\cite{Fishman_2023}. The fitted dip at $\tau=0$ gives $g^{(2)}(\tau=0)=0.25\pm 0.01$, below the classical limit of $g^{(2)}(\tau=0)=0.5$, affirming the presence of a single SnV center as opposed to an ensemble. Accounting for a $\sim$550~ps timing jitter of the detector (based on fit in Fig.~\ref{fig:purcell}), we extract the emitter's PSB signal to be $(4.38\pm 0.15)\times 10^3$~cps against a background of $290\pm 10$~cps~\cite{Fishman_2023}, which agrees with our measured detector dark counts. With background subtraction, we deduce $g^{(2)}(\tau=0)=0.14\pm 0.01$, limited by the detector jitter.

To ensure coupling between the SnV centers and the PhC cavity modes, we apply in-situ gas tuning with argon~\cite{Li_2015,Nguyen_2019_PRL,Rugar_2021} to overlap the cavity resonance $\lamcav$ with the SnV center's zero-phonon line (ZPL) $\lamsnv$ (see Supplement 1, Sec. 4). We resonantly excite the SnV centers with 500~ps-long pulses generated by an electro-optical modulator (see Supplement 1, Sec. 3). We verify emitter-cavity coupling by measuring the lifetime of SnV centers in each QMC channel.

Figure~\ref{fig:purcell} shows lifetime $\tau$ versus emitter-cavity detuning $\Delta=\lamcav-\lamsnv$ of a SnV center in cavity 6 in the QMC, with ZPL at 619.22~nm. Each lifetime curve is fitted with a single exponential convolved with a Gaussian that represents the instrument response function (IRF)~\cite{Stipcevic_2017} mainly set by the detector's timing resolution. At large detuning, i.e. $\Delta=7.45~$nm, the measured lifetime is $5.89\pm 0.25$~ns. As $\Delta$ decreases, the lifetime monotonically decreases. At close to zero detuning, i.e. $\Delta=0.05~$nm, the lifetime is shortened to $1.12\pm 0.04$~ns. The observed lifetime reduction signifies Purcell enhancement of the SnV center's spontaneous emission rate due to coherent emitter-cavity coupling~\cite{Purcell_1946}.

In order to estimate the maximum Purcell factor $F_P$ realized in experiments, we account for the modified local density of states when placing a SnV center in nanostructure and its non-radiative decay pathways out of the excited state by using the definition~\cite{Faraon_2011_natphoton,Li_2015}
\begin{align}
    F_P &= \frac{\tau_\text{bulk}}{\xi}\left(\frac{1}{\tau_\text{on}}-\frac{1}{\tau_\text{off}}\right),
\end{align}
where $\xi=0.456$ is a product of the SnV center's quantum efficiency $\text{QE}=0.80\pm 0.05$~\cite{Iwasaki_2017_PRL,Herrmann_2023} and the Debye-Waller factor $\text{DW}=0.57\pm 0.01$~\cite{Gorlitz_2020_NJP}. $\tau_\text{on}$ and $\tau_\text{off}$ are the (nearly-)on-resonant and off-resonant lifetimes. We also measure $\tau_\text{bulk}=5.10\pm 0.22~$ns typical of what is observed for SnV centers residing in bulk diamond. The estimated Purcell factor for the measured SnV center in Fig.~\ref{fig:purcell} is $F_P=8.07\pm 0.89$. Furthermore, we estimate the probability of the SnV center emitting into the cavity mode in the Purcell regime ($\kappa\gg\Gamma$) to be $\beta\approx F_P/(F_P+1)=89\pm 14\%$.

\begin{table}[h!]
	\begin{tabular}{>{\centering}p{1.5cm}>{}p{1.5cm}>{}p{2.3cm}>{}p{1.7cm}>{}p{1.5cm}>{}p{1.5cm}}
		\toprule
		\textbf{Channel} & \textbf{ZPL (nm)} & \textbf{Purcell Factor} & $\beta$ & $\tau_{\mathrm{off}}/\tau_{\mathrm{on}}$ & $\lambda_\text{cav,0}$ \textbf{(nm)}\\ \midrule
		2 & 619.2821 & 4.13$\pm$0.59 & 0.81$\pm$0.16 & 3.12$\pm$0.19 & 613.9\\ \midrule
		  4 & 619.2560 & 10.40$\pm$1.06 & 0.91$\pm$0.13 & 6.28$\pm$0.32 & 614.4\\ \midrule
        5 & 619.2965 & 5.32$\pm$0.72 & 0.84$\pm$0.16 & 3.65$\pm$0.23 & 615.1\\ \midrule
        6 & 619.2220 & 8.07$\pm$0.89 & 0.89$\pm$0.14 & 5.25$\pm$0.28 & 611.8\\
		\bottomrule
	\end{tabular}
	\caption{A summary table of four QMC channels exhibiting Purcell enhancement.}
    \label{tab:purcell}
\end{table}

Importantly, we achieve \textit{multi-channel} Purcell enhancement within a single QMC. We find four coupled emitter-cavity systems in channels 2, 4, 5 and 6 (channels 1 and 3 do not contain SnV centers within the cavities). Table~\ref{tab:purcell} shows a summary of each coupled emitter-cavity system's Purcell factor $F_P$, $\beta$-factor, and lifetime reduction ratios $\tau_\text{off}/\tau_\text{on}$. On average, the Purcell factors and $\beta$-factors acquired from this single QMC are $6.98\pm 0.42$ and $86\pm 7\%$, respectively.

In particular, one emitter-cavity coupled system in channel 4 exhibits ten-fold Purcell enhancement at $F_P=10.40\pm 1.06$. Its cavity $Q$ factor at close to resonance $\Delta=0.14~$nm is $Q\approx (2.28\pm 0.05)\times 10^3$ (see Supplement 1, Sec. 4). Accounting for detuning that reduces the measured Purcell factor by a factor of ${1+4Q^2(\lamsnv/\lamcav-1)}$~\cite{Faraon_2011_natphoton,Faraon_2012}, we estimate the maximum experimental Purcell factor at $\Delta=0$ improves slightly to $F_P=10.63\pm 0.16$. The theoretical maximum Purcell factor $F_{P,\text{max}}=\frac{3}{4\pi^2}\left(\frac{\lamcav}{n}\right)^3 \frac{Q}{V}$ is $F_{P,\text{max}}=216.2\pm 0.4$ (assuming the simulated mode volume). Accounting for dipole misalignment (angular difference between the [111] and [100] crystal axes) reduces the maximum Purcell factor to $F_{P,\text{max}}=124.9\pm 0.3$. We attribute the difference between the theoretical and experimental maxima of a factor of $F_{P,\text{max}}/F_P$$\sim$$11.75\pm 0.02$ to the SnV center being spatially off-centered from the cavity field maximum, an issue which can be resolved by employing focused ion beam implantation~\cite{Schroder_2017} or masked implantation~\cite{Nguyen_2019_PRB} in future efforts.

\section{Heterogeneous integration into a PIC}

\begin{figure}
    \centering
    \includegraphics[width=\textwidth]{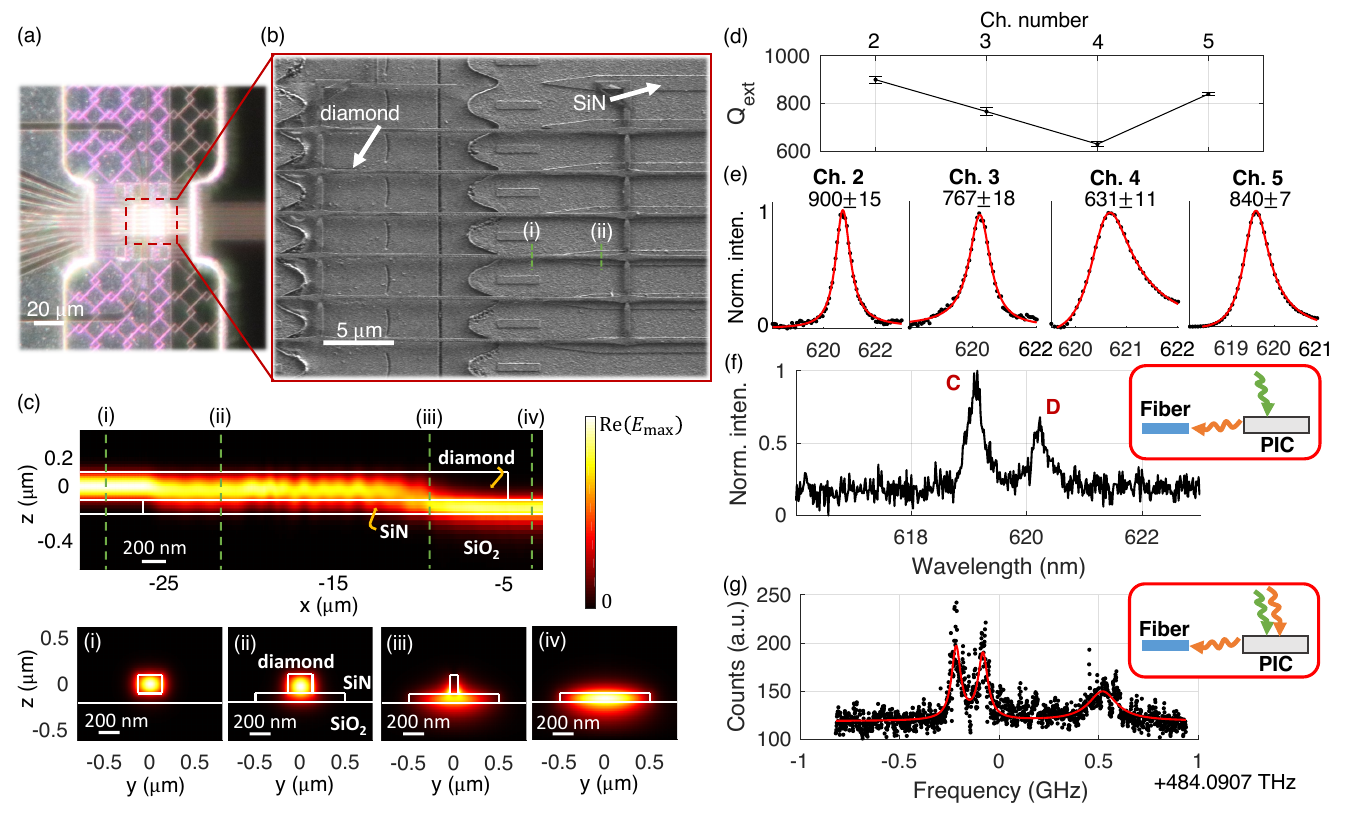}
    \caption{Integration of a diamond QMC into a SiN PIC. (a) An optical image of the integrated device. (b) A SEM image of the diamond QMC in the socket. (c) The TE mode propagates from diamond waveguide to the evanescently coupled SiN waveguide (100~nm thick) on oxide. The waveguide modes for the integrated QMC on SiN PIC: (i) diamond waveguide before contact with SiN, (ii) in the overlapped region at 10~$\upmu$m and (iii) at 2.25~$\upmu$m from the diamond tip, and (iv) SiN waveguide on oxide. Positions (i) and (ii) are also indicated in (b). (d) The fitted $Q$ factors of channels 2-5 based on the measured (e) cavity transmission spectra. Each cavity resonance is fitted with the Fano-Lorentz function (see Supplement 1, Sec. 2), with fitted $Q$ factors: 900$\pm$15, 767$\pm$18, 631$\pm$11, 840$\pm$7. (f) A PL spectrum using off-resonant 515~nm excitation reveals the inhomogeneously broadened C and D transitions of an ensemble of SnV centers in channel 2. (g) A PLE curve showing three peaks at 484.0904~THz, 484.0906~THz, and 484.0912~THz, with linewidths $55.2\pm 5.9~$MHz, $58.9763\pm 6.8~$MHz, $173.4969\pm 26.7~$MHz, respectively.}
    \label{fig:PIC_spectroscopy}
\end{figure}

To showcase hybrid integration with diamond color centers, we transfer-print a separate diamond QMC that is more preferentially waveguide-coupled (different from Fig.~\ref{fig:diamond_spectroscopy}(c)) into a silicon nitride PIC using a PDMS stamp (see Supplement 1, Sec. 6). The PIC supports independent low-loss waveguides coupled to all QMC channels with oxide-cladding designed for optimal edge-coupling with single-mode fibers~\cite{Sorace–Agaskar_2019,Starling_2023}. Identical to what is depicted in Fig.~\ref{fig:overview_architecture}, Fig.~\ref{fig:PIC_spectroscopy}(a) is a microscope image of the oxide opening, with white lines defining its boundaries. In dark field imaging, the ground and MW lines appear as the purple regions (with tessellated patterns on top). Extending into the oxide opening "socket" from both left and right sides are parallel SiN waveguides, on top of which we transfer-print the QMC~\cite{Chanana_2022,Raniwala_2023_CLEO}. Figure~\ref{fig:PIC_spectroscopy}(b) shows a SEM image of the socket containing the integrated QMC. Due to a pitch mismatch between the QMC and the SiN waveguides, only channels 2-5 are optically coupled for measurements. We estimate roughly a $\sim$21~$\upmu$m overlap in length between diamond and SiN waveguides, both of which are adiabatically tapered to minimize scattering loss. With the diamond (SiN) waveguide tapering down from 260~nm (1~$\upmu$m) to 50~nm (0~nm) in width over 9~$\upmu$m (4~$\upmu$m) in length, the transmission efficiency is simulated to be $94.4\%$ at 619~nm in FDTD at zero angular offset. Based on the SEM, we estimate an alignment angular offset of $\sim$$0.5$ degree with a corresponding transmission efficiency $\sim$$91\%$ (see Supplement 1, Sec. 6). Figure~\ref{fig:PIC_spectroscopy}(c) shows the propagating TE mode out of the cavity from the diamond waveguide (left) to the evanescently coupled SiN waveguide (right) on oxide, with mode profiles evaluated at four selected points: (i) suspended diamond waveguide before overlapping with SiN, (ii) in the overlapped region at 10~$\upmu$m and (iii) at 2.25~$\upmu$m from the diamond tip, and (iv) SiN waveguide on oxide. The SiN waveguide mode for each channel is then routed to an inversely tapered waveguide at the chip's edge for optimal optical coupling to a single-mode fiber~\cite{Starling_2023}.

Next, we perform cavity transmission measurements at room temperature. The efficient coupling between the diamond and SiN waveguides enables observation of cavity resonances by top excitation of the cavity mode and collection from an edge-coupled fiber. Figure~\ref{fig:PIC_spectroscopy}(d) displays the statistics of the measured $Q$ factors based on the Fano resonance spectra shown in Fig.~\ref{fig:PIC_spectroscopy}(e). The fitted $Q$ are 900$\pm$15, 767$\pm$18, 631$\pm$11, 840$\pm$7, respectively.

Furthermore, optical coupling between the diamond QMC and the PIC enables us to perform spectroscopy on SnV centers at 1.3~K. By exciting at the cavity center of channel 2 off-resonantly at 515~nm, we observe again C and D transitions of an ensemble of SnV centers as shown in Fig.~\ref{fig:PIC_spectroscopy}(f). We also perform PLE spectroscopy via resonant excitation to selectively probe individual SnV centers. Figure~\ref{fig:PIC_spectroscopy}(g) shows a PLE curve indicating the presence of three peaks, which could indicate either the presence of multiple emitters or nuclear-electro hyperfine transitions from the Sn-117 isotope (see Supplement 1, Sec. 10)~\cite{Harris_2023}. The fitted ZPL positions are 484.0904~THz, 484.0906~THz, and 484.0912~THz, with linewidths $\Gamma=55.2\pm 5.9~$MHz, $58.9763\pm 6.8~$MHz, $173.4969\pm 26.7~$MHz, respectively. The smallest observed linewidth is a factor of $\sim 2$ from the observed transform limit 27.8~MHz (based on the average detuned lifetime of the 4 devices studied in Table~\ref{tab:purcell}). Preservation of optical coherence of SnV centers showcases the suitability of the integrated QMC-PIC platform as an efficient spin-photon interface.

\section{Projected state transfer fidelity and success probability}

\begin{figure}
    \centering
    \includegraphics[width=0.7\textwidth]{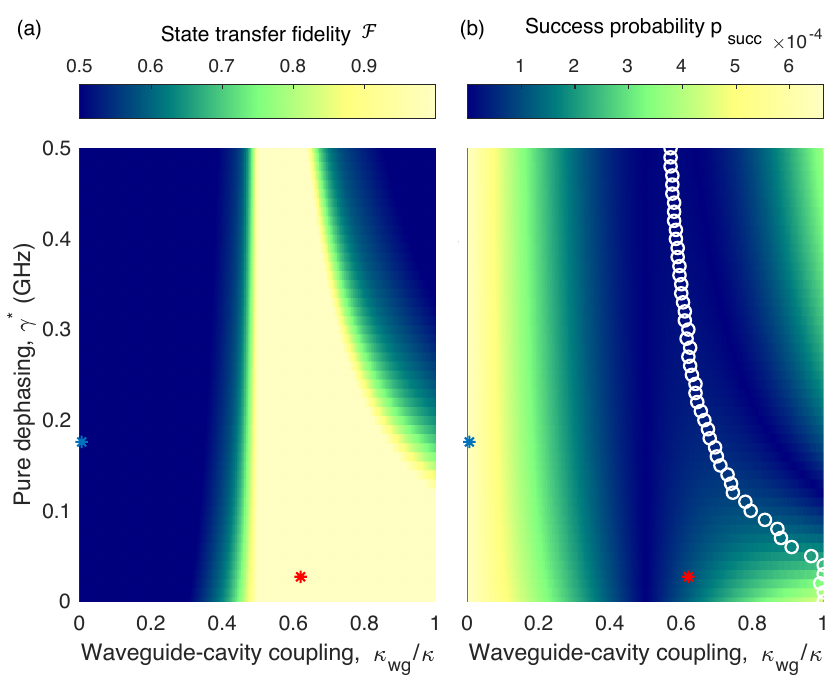}
    \caption{The photon-to-spin state transfer fidelity and success probability based on the cavity reflection protocol and experimental results. The (a) fidelity $\mathcal{F}$ and success probability $\psucc$ (b) are numerically computed as functions of both pure dephasing $\gamma^*$ and waveguide-cavity coupling $\kwg/\kappa$. The blue (red) dotted line indicates a cut at $\gamma^*=176~$MHz ($\gamma^*=27~$MHz) based on the linewidth in the 4~K (1.3~K) experiment. The blue star marker represents the demonstrated device, with $\kwg/\kappa=5\times 10^{-3}$ (based on FDTD) and $\gamma^*=176~$MHz. The red star marker indicates the performance with improvements in $\kwg/\kappa=0.62$~\cite{Knall_2022} and $\gamma^*=27~$MHz. The corresponding fidelity is unity and $\psucc\approx 10^{-4}$.}
    \label{fig:teleportation}
\end{figure}

The compactness of the PIC allows parallel addressing of multiple channels in the QMC, crucial for memory multiplexing that boosts the quantum state transfer rate between photonic and spin qubits. Importantly, optical routing in a miniaturized PIC provides intrinsic phase stability necessary for high-fidelity interference-based operations~\cite{Chen_2021}. In particular, we consider a cavity reflection-based protocol in which a polarization-encoded photonic qubit acquires a spin-state-dependent phase and is detected to herald photon-to-spin state transfer~\cite{Duan_2004,Tiecke_2014,Chen_2021} (see Supplement 1, Sec. 11).

In this protocol, the fidelity $\mathcal{F}$ hinges on the cavity reflection coefficient $r(\omega)$ (see Supplement 1, Sec. 11) as a function of pure dephasing $\gamma^*$, the emitter-cavity coupling strength $g$, and diamond waveguide-cavity coupling $\kwg/\kappa$. $\kwg$ is the cavity mode's decay rate into the waveguide and $\kappa\approx \kwg+\kappa_s$ is its total decay rate, where $\kappa_s$ is the scattering loss rate, as shown in Fig.~\ref{fig:overview_architecture}. Based on the cavity-enhancement results (Fig.~\ref{fig:diamond_spectroscopy}(f) and channel 4 in Fig.~\ref{fig:purcell}(b)), we estimate pure dephasing $\gamma^*=2\pi\times 176~$MHz and emitter-cavity coupling $g=\sqrt{\Gamma\kappa/4}\approx 2\pi\times 2.8~$GHz (see Supplement 1, Sec. 11). The investigated diamond cavities are designed to maximize $Q$ and are consequently undercoupled to the diamond waveguide mode. Based on FDTD simulations, the coupling strength is calculated to be $\kwg/\kappa$$\sim$$5\times 10^{-3}$.

Next, to estimate the success probability $\psucc$ of state transferring, we extract the relevant efficiencies by considering optical losses from the cavity center to the detector. Specifically, we analyze the three subsystems: (i) the device consisting of the diamond QMC and the SiN PIC, (ii) the fiber setup, and (iii) the free-space optics setup (see Supplement 1, Sec. 7). For (i), we account for waveguide-cavity coupling $\kwg/\kappa$, transmission efficiency from diamond waveguide to the SiN waveguide in the uncladded socket ($\sim$0.91 based on FDTD), directional coupler (0.5), transmission through the junction where uncladded and oxide-cladded regions meet (0.53~\cite{Starling_2023}), propagation in the PIC ($\sim$0.99~\cite{Sorace–Agaskar_2019}), and fiber edge-coupling efficiency (0.3 measured at 620~nm, see Supplement 1, Sec. 6). The estimated efficiency is $\eta_{(i)}\approx 3.2\times 10^{-4}$. For (ii), given the measured fiber transmission from inside the cryostat to the external setup ($\sim$0.61) and fiber insertion efficiency (conservative estimate of 0.89 per fiber-to-fiber connection), the estimated efficiency is $\eta_{(ii)}\approx 0.48$. Lastly, for (iii), the free-space setup's efficiency is $\eta_{(iii)}\approx 0.96$. Prior to detection, the overall collection efficiency is $\eta_{(i)}\cdot\eta_{(ii)}\cdot\eta_{(iii)}\approx 1.5\times 10^{-4}$. Assuming a detector's efficiency of 0.65, the overall detection efficiency is then $9.6\times 10^{-5}$, primarily limited by the waveguide-cavity coupling strength $\kwg/\kappa=$$\sim$$5\times 10^{-3}$.

Using our current device performance and treating the waveguide-cavity coupling $\kwg/\kappa$ as a variable, we estimate the fidelity and success probability that this quantum PIC platform could achieve for quantum state transfer, a key operation for quantum networking. Crucially, the protocol we consider is based on the Duan-Kimble scheme~\cite{Duan_2004} that relies on spin-state-dependent cavity reflection, in particular the $\ket{V}$ polarization mode reflects off the bare cavity whereas the $\ket{H}$ polarization mode acquires a phase from the spin coupled to a TE-mode cavity. We first extract an overall detection efficiency and excitation efficiencies $\etadet=1.9\times 10^{-2}$ and $\etaexc=3.4\times 10^{-2}$ \textit{without accounting for $\kwg/\kappa$}. The success probability is then defined as $\psucc=\etadet\etaexc |r|^2$, where $r=r(\omega,\kwg, \kappa)$ is the cavity reflection coefficient dependent on both the frequency and the waveguide-cavity coupling strength (see Supplement 1 Section 11. Sweeping over both $\kwg/\kappa$ and $\gamma^*$,  Fig.~\ref{fig:teleportation}(a,b) indicate a trade-off between $\mathcal{F}$ and $\psucc$ when pure dephasing is present.

For pure dephasing less than 0.5~MHz, $\mathcal{F}$ increases sharply close to the critical coupling regime $\kwg/\kappa=0.5$. Intuitively, in the undercoupling regime $\kwg/\kappa<0.5$, the phase contrast between $\rdown(\omega)$ and $\rup(\omega)$ is much less than $\pi$ (see Supplement 1, Sec. 11), hence limiting the spin-photon entanglement fidelity. As $\kwg/\kappa$ increases past $\kwg/\kappa=0.5$, $\mathcal{F}$ increases to unity and eventually rolls off with finite pure dephasing. The point at which $\mathcal{F}$ is maximized by an optimal choice of $\kwg/\kappa$ is marked as a white circle in Fig.~\ref{fig:teleportation}(b) for each value of $\gamma^*$. 

Based on the PLE curve shown in Fig.~\ref{fig:diamond_spectroscopy}(f), in which the total measured linewidth is $\Gamma=\gamma+\gamma^*$ where $\gamma\sim 1/\tau_\text{off}$ is the transform-limit, we estimate pure dephasing to be $\gamma^*=176~$MHz. Our current waveguide-cavity coupling is estimated to be $\kwg/\kappa=5\times 10^{-3}$, which leads to a fidelity at the classical limit $\mathcal{F}=0.5$, as shown by the blue marker in Fig.~\ref{fig:teleportation}(a). Next, we project how $\mathcal{F}$ and $\psucc$ vary with better optical coherence and improvements on the photonic cavity design. For the former, we use the narrowest linewidth out of the peaks displayed in Fig.~\ref{fig:PIC_spectroscopy}(g) with a corresponding pure dephasing of $\gamma^*=27~$MHz. As for the latter, we take state-of-the-art waveguide-cavity coupling of $\kwg/\kappa=0.62$ demonstrated in Ref.~\cite{Knall_2022}. Our calculation shows that the fidelity can improve to $\mathcal{F}\approx 1$ with $\psucc\approx 10^{-4}${, as shown by the red star markers in Fig.~\ref{fig:teleportation}(a,b).

Clearly, to boost both state transfer fidelity and success probability hinge critically on improving the waveguide-cavity coupling strength. As stated previously, our current device is only designed to maximize $Q$. One approach to better the photonic crystal cavity design is to adiabatically reduce the mirror strength of the waveguide-coupled side, entailing tapering down the air hole size and decreasing the number of holes~\cite{Knall_2022}. Another alternative is to utilize the ``sawfish`` photonic crystal cavity, which has no air holes but instead modulated waveguide width, a feature which is conducive to improve out-coupling to the adjacent waveguide~\cite{Chen_2021,Bopp_2022}. 

To further enhance success probability without trading off fidelity demands improving the total collection efficiency. First, the large scattering loss present at the oxide junction can be reduced to $0.5~$dB by employing inter-layer SiN waveguide coupling~\cite{Shang_2015}. Second, instead of using a fiber array with standard optical fiber, lensed fibers can be used to better mode matching and consequently the edge coupling efficiency. Additionally, using a detector with higher quantum efficiency, e.g. superconducting nanowire single-photon detectors, can boost the detection efficiency to 99$\%$~\cite{Bhaskar_2020}. As a result, the overall detection efficiency without considering diamond waveguide-cavity coupling can improve to 0.19, about an order of magnitude higher than the current value. 

Relatedly, improving various components contributing to the total collection efficiency is not only important for state transfer success probability, but also imperative for maintaining the SnV center's charge state during coherent quantum operations. Our current PLE and lifetime measurements require 515~nm illumination to stabilize the negatively-charged SnV centers~\cite{Gorlitz_2022}. As Ref.~\cite{Parker_2023} has shown, however, this need to apply aboveband light can be avoided by applying $<$~nW resonant laser power, which requires a high excitation/detection efficiency and may be within reach with the aforementioned ameliorating measures.

Lastly, reduction in pure dephasing also leads to eliminating the trade-off between fidelity and success probability, as shown by Fig.~\ref{fig:teleportation}(a,b). Although near-transform-limited linewidth has been observed in the SnV centers in the PIC-integrated QMC shown in Fig.~\ref{fig:PIC_spectroscopy}(g), there still remains defect centers exhibiting much worse optical dephasing such as the one presented in Fig.~\ref{fig:diamond_spectroscopy}(f). We suspect the variance in linewidths stems from the emitters' proximity to the sidewalls, which could introduce surface charge noises~\cite{OrphalKobin_2023}. Modifying the cavity design that reduce proximity to sidewalls~\cite{Mouradian_2018_phd,Bopp_2022} could then potentially lower $\gamma^*$ on top of benefiting waveguide-cavity coupling. Additional materials engineering may also be necessitated. Low-energy ion implantation to minimize lattice damage has been shown to improve optical coherence~\cite{Rugar_2020}. Studies of high-pressure and high-temperature (HPHT) treatment~\cite{Iwasaki_2017_PRL,Gorlitz_2020_NJP} have also demonstrated narrowing of the inhomogeneous distribution of defect centers' emissions in diamond. This could suggest HPHT heals the diamond lattice from implantation damage, and may minimize pure dephasing caused by nearby vacancy defects. Our current diamond samples have been implanted at relatively high dosage of $\sim$$10^{12}$ ions/cm$^2$. Future studies with lower implantation energy and subsequent HPHT treatment can potentially reduce pure dephasing due to defects formed from lattice damage.

\section{Outlook and conclusions}

In this work, we experimentally realize a multi-channel cavity-enhanced spin-photon interface in a single diamond QMC~\cite{Wan_2020}. We show $>50\%$ yield in demonstrating SnV-cavity coupling in four out of six channels, with an average Purcell factor of $\sim$7. We further demonstrate heterogeneous integration of diamond emitter-cavity systems into a SiN PIC. The photonic chip optically interposes to edge-coupled fibers that enable on-chip characterization of nanocavities with $Q$ exceeding $5\times 10^2$ and emitter spectroscopy unveiling an optical linewidth $\Gamma=66~$MHz close to the transform limit at 1.3~K. Based on the experimental results, we extract the relevant system parameters and simulate the photon-to-spin state transfer fidelity and success probability~\cite{Duan_2004,Tiecke_2014,Chen_2021}. Our findings suggest that this hybrid architecture can achieve spin-photon state transfer fidelity $\mathcal{F}$ nearing unity and success probability $\psucc$ exceeding $10^{-3}$ for building near-term quantum network nodes by considering improvements on several key components.

This work should motivate future efforts to use chip-integrated spin-cavity systems to perform multiplexed spin-photon entanglement and teleportation. This entails utilizing microwave lines to coherently control individual spin qubits~\cite{Christen_2023_CLEO} and integrating single photon detectors for more efficient optical measurements~\cite{Pernice_2012,Najafi_2015}. The integration approach can also be applied to active PIC platforms not covered in this work, such as aluminum nitride~\cite{Soltani_2016_AlN,Lu_2018_AlN,Wan_2020,Dong_2022}, gallium phosphide~\cite{Fu_2008,Wilson_2019}, lithium niobate~\cite{Desiatov_2019,Descamps_2023}, and silicon carbide~\cite{Lukin_2020,Anderson_2022}, all of which permit high-bandwidth switching networks for high entanglement generation rates, where memory multiplexing eliminates the effect of memory saturation~\cite{van_Dam_2017,Shchukin_2019,Lee_2020,Dai_2021,Dhara_2022_PRA,Choi_2023,Chen_2023_satellite}. The discussed physics and engineering platform extend beyond the use of color centers in diamond. Other promising solid-state qubits such as quantum dots~\cite{Laucht_2012,Lodahl_2018,Sun_2018}, color centers in silicon carbide~\cite{Lukin_2020}, rare-earth ions~\cite{Zhong_2015,Raha_2020}, and defects in silicon~\cite{Higginbottom_2022,DeAbreu_2023,Prabhu_2023,Saggio_2023} can benefit as well from the many functionalities bestowed by integrated photonics. Lastly, directional couplers form phase-stable on-chip interferometers for entangling spin-cavities between different local channels, allowing building cluster states crucial for measurement-based quantum computation~\cite{Raussendorf_2003} and error-corrected quantum communication~\cite{Borregaard_2020}.

\section{Backmatter}

\begin{backmatter}
\bmsection{Funding}

\bmsection{Acknowledgments}
We thank Isaac Harris, Eric Bersin, and Madison Sutula for insightful discussions. K.C.C. acknowledges funding support by the National Science Foundation (NSF) RAISE-TAQS 1839155 and the MITRE Corporation Moonshot program. I.C. acknowledges support from the National Defense Science and Engineering Graduate (NDSEG) Fellowship Program, the NSF EFRI ACQUIRE program EFMA-1641064 and NSF award DMR-1747426. H.R. acknowledges support from the NDSEG Fellowship Program and the NSF Center for Ultracold Atoms (CUA) No. 1734011. L.D. acknowledge funding from the European Union’s Horizon 2020 research and innovation program under the Marie Sklodowska-Curie grant agreement No. 840393. L.L. acknowledges funding from NSF QISE-NET award DMR-1747426 and the ARO MURI W911NF2110325. M.T. performed this work, in part, with funding from the Center for Integrated Nanotechnologies, an Office of Science User Facility operated for the US Department of Energy Office of Science. D.E. acknowledges funding support from the NSF RAISE TAQS program 1839155, the NSF Engineering Research Center for Quantum Networks (CQN), awarded under cooperative agreement number 1941583, and the NSF C-ACCEL program.

\bmsection{Disclosures}
Distribution Statement A. Approved for public release. Distribution is unlimited. This material is based upon work supported by the National Reconnaissance Office and the Under Secretary of Defense for Research and Engineering under Air Force Contract No. FA8702-15-D-0001. Any opinions, findings, conclusions or recommendations expressed in this material are those of the authors and do not necessarily reflect the views of the National Reconnaissance Office or the Under Secretary of Defense for Research and Engineering. $\copyright$ 2023 Massachusetts Institute of Technology. Delivered to the U.S. Government with Unlimited Rights, as defined in DFARS Part 252.227-7013 or 7014 (Feb 2014). Notwithstanding any copyright notice, U.S. Government rights in this work are defined by DFARS 252.227-7013 or DFARS 252.227-7014 as detailed above. Use of this work other than as specifically authorized by the U.S. Government may violate any copyrights that exist in this work.

D.E. holds shares in Quantum Network Technologies; these interests have been disclosed to MIT and reviewed in accordance with its conflict of interest policies, with any conflicts of interest to be managed accordingly.

\bmsection{Data Availability Statement}
All relevant experimental data are available within the Article and its Supplementary Information. Numerical simulation data are available on request from authors.

\bmsection{Supplemental document}
See Supplement 1 for supporting content.

\end{backmatter}

\bibliography{references_cavityPIC}

\end{document}